\documentclass[fleqn,twoside,twocolumn,nofootinbib,showkeys]{revtex4} 
\usepackage[nocpr]{ujp} 
\begin{document}
\title[Hartree--Fock Problem of an Electron-Hole
Pair]
{HARTREE--FOCK PROBLEM OF AN ELECTRON-HOLE PAIR IN THE QUANTUM WELL
GaN}%
\author{L.E. Lokot}
\affiliation{V.E. Lashkaryov Institute of Semiconductor Physics, Nat.
Acad. of Sci. of Ukraine}
\address{41, Prosp. Nauky, Kyiv 03028, Ukraine}
\email{llokot@gmail.com}
\udk{533.9} \pacs{73.21.Fg, 78.45.+h,\\[-3pt] 78.47.da, 42.55.Px}
\razd{\secix}


 \autorcol{L.E.\hspace*{0.7mm}Lokot}

\setcounter{page}{56}%

\begin{abstract}
We present microscopic calculations of the absorption spectra for
$\textrm{GaN}/\textrm{Al}_{x}\textrm{Ga}_{1-x}\textrm{N}$ quantum
well systems. Whereas the quantum well structures with the parabolic
law of dispersion exhibit the usual bleaching of an exciton
resonance without shifting a spectral position, the significant
red-shift of an exciton peak is found with increasing the
electron-hole gas density for a wurtzite quantum well. The energy of
the exciton resonance for a wurtzite quantum well is found. The
obtained results can be explained by the influence of the valence
band structure on quantum confinement effects. The optical gain
spectrum in the Hartree--Fock approximation and the Sommerfeld
enhancement are calculated. A red shift of the gain spectrum in the
Hartree--Fock approximation with respect to the Hartree gain
spectrum is found.
\end{abstract}

\maketitle

\section{Introduction}

The physical properties of wide bandgap group-III quantum well
systems are under investigation due to their application to light
emitters and semiconductor lasers in the ultraviolet, blue, and green
wavelength regions. Ultraviolet light-emitting diodes and lasers have
recently obtained considerations due to applications to the compact
biological detection systems, analytical devices, and medical
diagnostics. A number of light-emitting diodes and laser diodes have
been demonstrated~\cite{{Savage},{Khan2}}. However, these  structures
are in the developmental stage, and there are many questions with
respect to the performance and device configurations.

Realizing the deep-ultraviolet semiconductor-based light-emitting
diodes provides light sources for various applications, for instance
to the biological detection and the data storage. Although such
devices basically need a
$\textrm{Al}_{x}\textrm{Ga}_{1-x}\textrm{N}$-based quantum well with
high Al contents, their fundamental optical properties remain under
discussion. It has been proved experimentally that the surface
emission from [0001]-oriented
$\textrm{Al}_{x}\textrm{Ga}_{1-x}\textrm{N}$ is quite weak because
of the predominant optical polarization along the [0001] $c$
direction \cite{{Kawanishi1},{Kawanishi2},{Shakya}}. The explanation
of these effects may be found from the difference of structures of
the valence bands in AlN and in GaN. In wurtzite GaN or AlN, the
degeneracy of the $p$-like states at the $\Gamma$ point is lifted by
both crystal-field splitting and spin-orbit splitting leading to
forming three valence bands at the Brillouin zone
center.

Since AlN has a negative crystal field splitting energy, while GaN
has a positive one, these splittings lead to the ordering of the valence
band in AlN: $\Gamma_7$, $\Gamma_9,$ and $\Gamma_7.$
Whereas we have $\Gamma_9$, $\Gamma_7,$ and $\Gamma_7$ in GaN \cite{{Banal}}.
Therefore, the topmost of the valence band in AlN has the crystal field
split off holes with $p_{z}$-states, while the topmost in GaN has
the heavy holes with $p_{x}$-like and $p_{y}$-like states, where the
axis $z$ is directed along the hexagonal axis.

Therefore, the emission from
$\textrm{Al}_{x}\textrm{Ga}_{1-x}\textrm{N}$ with high (low) Al-content is polarized along
(perpendicular to) the $c$ axis.

Recently, many studies have been focused on the potential
application of nanostructures, such as photonic crystal structures,
nanoholes, nanodots, and nanorods. In the studies of the technology
involving the photonic band gap, it seems that, in the case of
dielectric rod nanoarrays or nanocolumns, a large gap is opened for
the TM mode, but not for the TE one~\cite{{Meade}}. Thus, with this
type of structures for laser applications, the light source in the
TM mode is \mbox{obtained.}\looseness=1

In the $c$-plane of InGaN/GaN quantum well systems, the compressive
strain is induced in the active layer, and the
light is TE-polarized~\cite{{Park}}. Furthemore,
there is a strong internal electric field caused by the spontaneous
and piezoelectric polarization charges at the interfaces of the
$c$-plane of the InGaN/GaN quantum well. This phenomenon leads to the
quantum confined Stark effect, decreases the internal quantum
efficiency, and leads to the emission spectrum which is red-shifted.

In some studies~\cite{{Schubert},{Kim},{Park1}} of interface
polarization charges, alloy materials were used to make a better
performance. Many works have focused on the nonpolar and semipolar
planes~\cite{{Romanov},{Yamaguchi},{Huang},{Nido}}. These results
have testified that the light emission will be polarized, and the
quantum confined Stark effect will be reduced. However, due to a
higher cost of the $a$- and $m$-plane substrates, it would be better
to use the $c$-plane substrate. In work ~\cite{{Chichibu},{Fu}}, the
$c$-plane of the InGaN/AlGaN quantum well structure was considered
instead of that of InGaN/GaN in order to obtain a tensile strain in
the quantum well layer. The previous studies and calculations have
shown that the $|Z\rangle$-like state is generated in nitride
materials, if the quantum well layer is under a tensile biaxial
strain.\looseness=1

Besides the nitride-based devices, the group-II oxides have been
considered for highly efficient laser
diodes~\cite{{Dang},{Fujita}} and high-performance field-effect
transistors~\cite{{Sasa},{Koike}}. The induced piezoelectric field
plays a significant role for both band structure and optical
gain~\cite{{Park2}}. However, the orientation of a crystal
structure significantly modifies the band structure through the strain
effect ~\cite{{Willatzen}}. It has been proved experimentally that
the growth along crystal directions different from the [0001] direction
leads to an increase in the quantum efficiency by decreasing the
strain-induced electric field in the quantum well region, possibly
leading to the ways of obtaining highly efficient white laser
diodes~\cite{{Auld}}. There are the theoretical works studying the
effects of crystal orientation on the piezoelectric field in
a strained wurtzite quantum well~\cite{{Willatzen},{Duggen}}. However,
the piezoelectric effect consists not only of a strain-induced
polarization; it also takes the response of both
electric field and polarization on the strain into consideration. These effects were
studied in paper~\cite{{Duggen}}.

A deeper understanding of the influence of band structures on
optical properties should help one to answer many questions. In
addition, the interesting effects of strong electron-hole Coulomb
interaction are presented in these materials. Many-body interactions
lead to effects, which consist the screening, dephasing, bandgap
renormalization, and phase-space filling
\mbox{\cite{{Lindberg},{Chowbook},{Chow},{Chow1}}.}

A general phenomenon of Coulomb enhancement may be explained as
follows. Due to the Coulomb attraction, an electron and a hole
have a larger tendency to be located near each other, than that in
the case of noninteracting particles. This increase of the
interaction duration leads to an increase of the optical transition
probability.

The paper is organized as follows. In Section~2, we present the
microscopic many-body theory, which is based on the Bloch equations
for semiconductors, i.e., the Heisenberg equations for the optical
polarization and the populations of carriers. In Section 3, we
consider a quantum well, which is oriented perpendicularly to the
growth direction [0001]. We research the overlap integral of
electron and hole wave functions and calculate the exciton binding
energy in the quantum well. We calculate the Hartree and
Hartree--Fock gain spectra. We calculate the exciton absorption
spectra in the wurtzite quantum well and compare them with the
absorption spectra in a quantum well with parabolic bands. We
calculate the Hartree and Hartree--Fock renormalization energy
spectra and the red shift of the gain spectra caused by an
electron-electron and hole-hole Coulomb interaction. A significant
Sommerfeld enhancement of the spectrum is determined. This
enhancement of the electric dipole moment caused by the
electron-hole Coulomb att\-raction.

\section{Theory}

Let us consider the points of zero slope, i.e., the points at which
the speed components $\frac{\partial\,E}{\partial\,k_{\alpha}}$ are
identically equal zero according to the symmetry conditions taking
into account the time inversion invariance. These points are
determined by the formula
$N=\frac{1}{h'}\sum_{g\in\,G}\frac{1}{2}[\chi_{v}^{2}(g)+\chi_{v}(g^{2})]\frac{1}{2}[\chi_{\psi}^{2}(g)+\chi_{\psi}(g^{2})]$.
In this case, all momentum components become zero, i.e.,
$\frac{\partial\,E}{\partial\,k_{\alpha}}=0$ for an all
directions~\textbf{k}~\cite{{Bir}}.

We consider a quantum well, which is oriented perpendicularly to the
growth direction [0001]. The axis $z$ is directed along the hexagonal $c$
axis. Then a longitudinal wave vector $k_{z}$ is changed by the
operator $k_{z}\rightarrow\,-i\frac{\partial}{\partial\,z}$. From the
Schr\"{o}dinger equation, we obtain the energy spectrum
$E_{n}(k_{t})$ for holes and for electrons, where
$k_{t}=(k_{x},k_{y})$ is a transversal wave vector. The necessary
condition of a band extremum in a vicinity of the band gap is the zero
derivative of the energy with respect to $k_{t}$. It is known from
semiconductor physics that the absorption spectrum in a vicinity
of the band gap with regard for a coupled electron-hole pair leads to an
exciton spectrum. The excitons mathematically obey the Schr\"{o}dinger
equation for a hydrogen atom, which is known as the Wannier
equation~\cite{{Noks}}.

The complete orthonormal system of functions for holes depends on
three quantum numbers: $\alpha$ that defines the number of a
subband, $\textbf{p}$ -- quasimomentum, and $m$ -- the number of
terms in the expansion of a wave function in the complete
orthonormal system of functions on the interval [$-w/2$, $w/2$],
which defines the width $w$ of the quantum well (see
works~\cite{{Lokot},{Lokot1}}). For electrons, the number, which
defines the number of a term in the expansion, is equal of the
number, which defines the number of a subband. In the paper, one
lowest conduction subband and one highest valence subband are
considered. In the electron-hole representation, we introduce the
operators of creation and annihilation for electrons and holes
$\hat{a}_{\textbf{p}}$, $\hat{a}_{\textbf{p}}^{+}$,
$\hat{b}_{-\textbf{p}}$, and $\hat{b}_{-\textbf{p}}^{+}$, where
$\textbf{p}=(p_{x},p_{y})$ is the transversal quasimomentum of
carriers in the plane of a quantum well. There is no necessity in
the quantum number, which defines the number of a subband.
Consequently, for a heavy hole, we have
\begin{equation}
\Psi=\sum_{\textbf{p}}\hat{b}_{\textbf{p}}\psi_{\textbf{p}}(\textbf{r}),
\end{equation}
where\vspace*{-3mm}
\begin{equation}
\psi_{\textbf{p}}(\textbf{r})=\frac{e^{i\,{\bf p}\,{\boldsymbol
{\rho}}}}{\sqrt{A}}\,|\textbf{p}\rangle,
\end{equation}
and $A$ is the area of a quantum well in the $(x,y)$ plane;
\begin{equation}
|\textbf{p}\rangle=\left\|
\begin{array}{cccc}
\phi_{\alpha}^{(1)}(z,\textbf{p})\\
\phi_{\alpha}^{(2)}(z,\textbf{p})\\
\phi_{\alpha}^{(3)}(z,\textbf{p})\\
\end{array}
\right\|,\end{equation}\vspace*{-5mm}
\begin{equation}
\phi_{\alpha}^{(j)}=\sum_{i=1}^{n}V_{\textbf{p}}^{(j)}[i,\alpha]\,\chi_{i}(z),
\end{equation}\vspace*{-5mm}
\begin{equation}
\chi_{n}(z)=\sqrt{\frac{2}{w}}\,\sin{\left(\!\pi\,n\,\left(\!\frac{z}{w}+\frac{1}{2}\!\right)\!\right)}\!,
\end{equation}
where $n$ is a natural number, $\alpha$='heavy hole'. For an
electron,
\begin{equation}
\Psi=\sum_{\textbf{p}}\hat{a}_{\textbf{p}}\psi_{\textbf{p}}(\textbf{r}),
\end{equation}
where\vspace*{-3mm}
\begin{equation}
\psi_{\textbf{p}}(\textbf{r})=\frac{e^{i\,{\bf p}\,{\boldsymbol
{\rho}}}}{\sqrt{A}}\chi_{1}(z).
\end{equation}

To make the analysis as simple as possible, we assume a nondegenerate
situation described by the Hamiltonian
$\hat{H}=\hat{H}_{0}+\hat{V}+\hat{H}_{\rm int}$, which is composed
of the kinetic energy of an
electron $\epsilon_{e,\textbf{p}}^{\nu_{e}}$ and the kinetic energy
of a hole $\epsilon_{h,\textbf{p}}^{\nu_{h}}$ in the electron-hole representation:
\begin{equation}
\hat{H}_{0}=\sum_{\textbf{p}}{\epsilon_{e,\textbf{p}}^{\nu_{e}}\hat{a}_{\textbf{p}}^{+}\hat{a}_{\textbf{p}}+\epsilon_{h,\textbf{p}}^{\nu_{h}}\hat{b}_{-\textbf{p}}^{+}\hat{b}_{-\textbf{p}}},
\end{equation}
where \textbf{p} is the transversal quasimomentum of carriers in the
plane of the quantum well, $\hat{a}_{\textbf{p}}$,
$\hat{a}_{\textbf{p}}^{+}$, $\hat{b}_{-\textbf{p}}$, and
$\hat{b}_{-\textbf{p}}^{+}$ are the annihilation and creation
operators of an electron and a hole. The Coulomb interaction
Hamiltonian for particles in the electron-hole representation takes
the form:
\[
\hat{V}=\frac{1}{2}\sum_{\textbf{p},\textbf{k},\textbf{q}}V_{q}^{\nu_{e}\nu_{e}\nu_{e}\nu_{e}}\hat{a}_{\textbf{p}+\textbf{q}}^{+}\hat{a}_{\textbf{k}-\textbf{q}}^{+}\hat{a}_{\textbf{k}}\hat{a}_{\textbf{p}}+\]\vspace*{-5mm}
\[
+V_{q}^{\nu_{h}\nu_{h}\nu_{h}\nu_{h}}\hat{b}_{\textbf{p}+\textbf{q}}^{+}\hat{b}_{\textbf{k}-\textbf{q}}^{+}\hat{b}_{\textbf{k}}\hat{b}_{\textbf{p}}-\]\vspace*{-5mm}
\begin{equation}-2\,V_{q}^{\nu_{e}\nu_{h}\nu_{h}\nu_{e}}\hat{a}_{\textbf{p}+\textbf{q}}^{+}\hat{b}_{\textbf{k}-\textbf{q}}^{+}\hat{b}_{\textbf{k}}\hat{a}_{\textbf{p}},
\end{equation}
where\vspace*{-3mm}
\[
V_{q}^{\nu_{\alpha}\nu_{\beta}\nu_{\beta}\nu_{\alpha}}\!=\!\frac{e^{2}}{\varepsilon}\frac{1}{A}\int\limits_{-w/2}^{+w/2}dz\int\limits_{-w/2}^{+w/2}dz'\chi_{\nu_{\alpha}}(z)\chi_{\nu_{\beta}}(z')\frac{2\pi}{q}\times\]\vspace*{-5mm}
\begin{equation}
\times\,e^{-q|z-z'|}\chi_{\nu_{\beta}}(z')\chi_{\nu_{\alpha}}(z),
\end{equation}
is the Coulomb potential of the quantum well, $\varepsilon$ is the
dielectric permittivity of a host material of the quantum well, and
$A$ is the area of the quantum well in the $xy$ plane.

The Hamiltonian of the interaction of a dipole with an
electromagnetic field is described as follows:
\[
\hat{H}_{\rm
int}=-\frac{1}{A}\sum_{\nu_{e},\nu_{h},\textbf{p}}((\mu_{\textbf{p}}^{\nu_{e}\nu_{h}})^{\star}\hat{p}_{\textbf{p}}^{\nu_{e}\nu_{h}}E^{\star}e^{i\omega\,t}+\]\vspace*{-5mm}
\begin{equation}
+(\mu_{\textbf{p}}^{\nu_{e}\nu_{h}})(\hat{p}_{\textbf{p}}^{\nu_{e}\nu_{h}})^{+}Ee^{-i\omega\,t}),
\end{equation}
where
$\hat{p}_{\textbf{p}}^{\nu_{e}\nu_{h}}=\langle\,\hat{b}_{-\textbf{p}}\hat{a}_{\textbf{p}}\rangle$
 is a microscopic dipole due to an electron-hole pair with the electron
(hole) momentum \textbf{p} (--\textbf{p}) and the subband number
$\nu_{e}$ ($\nu_{h}$),
$\mu_{\textbf{k}}^{\nu_{e}\nu_{h}}=\int{d^{3}rU_{j'\sigma'\,\textbf{k}}\textbf{e}\hat{\mathbf{p}}U_{j\sigma\,\textbf{k}}}$,
 is the matrix element of the electric dipole moment, which depends
on the wave vector \textbf{k} and the numbers of subbands, between which
the direct interband transitions occur, $\textbf{e}$ is a unit vector
of the vector potential of an electromagnetic wave,
$\hat{\mathbf{p}}$ is the momentum operator. Subbands are described
by the wave functions $U_{j'\sigma'\,\textbf{k}}$,
$U_{j\sigma\,\textbf{k}}$, where $j'$  is the number of a subband from
the conduction band, $\sigma'$ is the electron spin, $j$ is the number
of a subband from the valence band, and $\sigma$ is the hole spin. We
consider one lowest conduction subband $j'=1$ and one
highest valence subband $j=1$. $E$ and $\omega$  are the electric
field amplitude and frequency of an optical wave.

We accept the approximation which simplifies the calculations in solving
the problem concerning the electron-hole gas. Namely, we consider the problem in
the case of a high density of the electron-hole gas (case $r_{s}<1$).
Estimating the ratio of the Coulomb potential energy to the Fermi energy,
we obtain
\begin{equation}
r_{s}=\frac{E_{\rm C}}{E_{\rm
F}}=\frac{2me^{2}}{\varepsilon\,\hbar^{2}\sqrt{n}\pi}=0.73
\end{equation}
for the concentration of the electron-hole gas $n=$ $=10^{13}$
cm$^{-2}$, the dielectric permittivity $\varepsilon=9.38$, the
transversal effective mass of an electron at $\Gamma$ point $m=0.18$
(inverse second derivative of the energy with respect to the
transversal wave vector). This indicates that the Fermi energy
dominates relative to the Coulomb potential energy as
$r_{s}\rightarrow\,0$ and increases more rapidly than the Coulomb
energy with the increasing density. As $r_{s}\rightarrow\,0$ the
terms corresponding to cyclic diagrams will dominate.

The Heisenberg equation for the electron,
$\hat{n}_{\textbf{p}}^{\nu_{e}}=$
$=\langle\,\hat{a}_{\textbf{p}}^{+}\hat{a}_{\textbf{p}}\rangle$, and
hole,
$\hat{n}_{\textbf{p}}^{\nu_{h}}=\langle\,\hat{b}_{-\textbf{p}}^{+}\hat{b}_{-\textbf{p}}\rangle$,
populations is written in the form:
\begin{equation}
\frac{\partial\,\hat{n}_{\textbf{p}}^{\nu_{e}}}{\partial\,t}=\frac{i}{\hbar}[\hat{H},\hat{n}_{\textbf{p}}^{\nu_{e}}].
\end{equation}
Substituting (8), (9), and (11) in (13), we obtain
\[
\hbar\,\frac{\partial\,\hat{n}_{\textbf{p}}^{\nu_{e}}}{\partial\,t}=-2{\rm
Im}[\mu_{\textbf{p}}^{\nu_{e}\nu_{h}}E(t)(\hat{p}_{\textbf{p}}^{\nu_{e}\nu_{h}})^{\star}]+i\sum_{\textbf{k}',\textbf{q}}V(q)\times\]\vspace*{-5mm}
\[\times(\langle\,\hat{a}_{\textbf{p}}^{+}\hat{a}_{\textbf{k}-\textbf{q}}^{+}\hat{a}_{\textbf{p}-\textbf{q}}\hat{a}_{\textbf{k}}\rangle-\langle\,\hat{a}_{\textbf{p}+\textbf{q}}^{+}\hat{a}_{\textbf{k}-\textbf{q}}^{+}\hat{a}_{\textbf{p}}\hat{a}_{\textbf{k}}\rangle +\]\vspace*{-5mm}
\begin{equation}
+\langle\,\hat{a}_{\textbf{p}}^{+}\hat{b}_{\textbf{k}-\textbf{q}}^{+}\hat{b}_{\textbf{k}}\hat{a}_{\textbf{p}-\textbf{q}}\rangle-\langle\,\hat{a}_{\textbf{p}+\textbf{q}}^{+}\hat{b}_{\textbf{k}-\textbf{q}}^{+}\hat{b}_{\textbf{k}}\hat{a}_{\textbf{p}}\rangle).
\end{equation}
Factorizing the convolutions of operators with the help of the Wick
theorem, the Heisenberg equation for an electron population reads
\begin{equation}
\hbar\,\frac{\partial\,\hat{n}_{\textbf{p}}^{\nu_{e}}}{\partial\,t}=-2{\rm
Im}[[\mu_{\textbf{p}}^{\nu_{e}\nu_{h}}E(t)+\sum_{\textbf{q}}V(q)\hat{p}_{\textbf{p}+\textbf{q}}^{\nu_{e}\nu_{h}}](\hat{p}_{\textbf{p}}^{\nu_{e}\nu_{h}})^{\star}].
\end{equation}
The pairwise convolutions originate from the $\psi$ operators, which
are taken at different points (this is the Hartree--Fock
approximation).

In the second order, the Coulomb potential energy reads
\[
\hbar\,\frac{\partial\,\hat{n}_{\textbf{p},{\rm
scat}}^{\nu_{e}}}{\partial\,t}=-\sum_{\textbf{k},\textbf{q}}2\pi\,V^{2}(q)\times\]\vspace*{-5mm}
\[\times\,D(\epsilon_{e}(\textbf{p})+\epsilon_{e}(\textbf{k}+\textbf{q})-\epsilon_{e}(\textbf{k})-\epsilon_{e}(\textbf{p}+\textbf{q}))\times\]\vspace*{-5mm}
\[\times[\hat{n}_{\textbf{p}}^{\nu_{e}}\hat{n}_{\textbf{k}+\textbf{q}}^{\nu_{e}}(1-\hat{n}_{\textbf{k}}^{\nu_{e}})(1-\hat{n}_{\textbf{p}+\textbf{q}}^{\nu_{e}})-\]\vspace*{-5mm}
\[-(1-\hat{n}_{\textbf{p}}^{\nu_{e}})(1-\hat{n}_{\textbf{k}+\textbf{q}}^{\nu_{e}})\hat{n}_{\textbf{k}}^{\nu_{e}}\hat{n}_{\textbf{p}+\textbf{q}}^{\nu_{e}}]-\]\vspace*{-5mm}
\[-\sum_{\textbf{k},\textbf{q}}2\pi\,V^{2}(q)D(\epsilon_{e}(\textbf{p})+\epsilon_{h}(\textbf{k})-\epsilon_{e}(\textbf{p}+\textbf{q})-\epsilon_{h}(\textbf{k}+\textbf{q}))\times\]\vspace*{-5mm}
\[\vspace*{-5mm}
\times[\hat{n}_{\textbf{p}}^{\nu_{e}}\hat{n}_{\textbf{k}}^{\nu_{h}}(1-\hat{n}_{\textbf{p}+\textbf{q}}^{\nu_{e}})(1-\hat{n}_{\textbf{k}+\textbf{q}}^{\nu_{h}})-\]\vspace*{-5mm}
\begin{equation}
-(1-\hat{n}_{\textbf{p}}^{\nu_{e}})(1-\hat{n}_{\textbf{k}}^{\nu_{h}})\hat{n}_{\textbf{p}+\textbf{q}}^{\nu_{e}}\hat{n}_{\textbf{k}+\textbf{q}}^{\nu_{h}}],
\end{equation}
where $D(\Delta)=\delta(\Delta)-i\pi^{-1}P(\Delta)$, and $P$ denotes
principal value.

We assume that
\begin{equation}
\frac{\partial\,\hat{n}_{\textbf{p}}^{\nu_{e}}}{\partial\,t}=\frac{\partial\,\hat{n}_{\textbf{p}}^{\nu_{h}}}{\partial\,t}=0.
\end{equation}
One can find the expectation value from the convolution of two
operators: $\langle\hat{a}_{k}^{+}\hat{a}_{p}(\tau)\rangle$
regarding the density matrix, i.e. the certain statistic operator
$\rho=\frac{e^{-\beta\,H_{0}}}{{\rm Sp}({e^{-\beta\,H_{0}}})}$. From
the Heisenberg equation, we obtain
\begin{equation}
\langle\hat{a}_{p}\hat{a}_{k}^{+}\rangle=e^{\beta\,\epsilon_{p}^{\nu_{e}}}\langle\hat{a}_{k}^{+}\hat{a}_{p}\rangle.
\end{equation}
Since
$\hat{a}_{p}\hat{a}_{k}^{+}=\delta_{pk}-\hat{a}_{k}^{+}\hat{a}_{p}$
for fermions, Eq. (18) yields the expression for the electron
population in terms of the Fermi distribution function:
\begin{equation}
\langle\hat{a}_{k}^{+}\hat{a}_{p}\rangle=\frac{\delta_{pk}}{1+e^{\beta\epsilon_{p}^{\nu_{e}}}},
\end{equation}
where $\epsilon_{p}^{\nu_{e}}=\varepsilon_{p}^{\nu_{e}}-E_{\rm  F}$,
and $E_{\rm  F}$ is the Fermi energy.

To calculate the sum in the ground-state energy of the
electron gas in all orders of perturbation theory,
the propagator is taken as a function \cite{{Gell-Mann}},
whose Fourier transform is equal to
\begin{equation}
Q_{q}(u)=\int\,d^{3}p\int\limits_{-\infty}^{\infty}e^{ituq}e^{-|t|[\frac{1}{2}q^{2}+\textbf{qp}]}dt.
\end{equation}
Works~\cite{{Gell-Mann},{Kittel},{Mattuck}} gave the direct
correspondence between the diagrams of the given order and the
integrals, whose Fourier transformations are
\begin{equation}
A_{n}=\frac{q}{2\pi\,n}\int\limits_{-\infty}^{\infty}du[Q_{q}(u)]^{n}.
\end{equation}

The complete contribution of all cyclic diagrams in the $n$-order of
perturbation theory is shown
\cite{{Gell-Mann},{Kittel},{Mattuck}} to be
\[\epsilon'\equiv\,\epsilon^{(2)}+\epsilon^{(3)}+\epsilon^{(4)}+...=\]\vspace*{-5mm}
\[=-\frac{3}{8\pi^{5}}\int\,\frac{d^{3}q}{q^{3}}\frac{1}{2\pi}\sum_{n=2}^{\infty}\langle\,[[\hat{F},\underbrace{\hat{V}],...\hat{V}}_{n-1}]\rangle\times\]\vspace*{-5mm}
\[\times\int\limits_{-\infty}^{\infty}du\frac{(-1)^{n}}{n}[Q_{q}(u)]^{n}(\frac{\alpha\,r_{s}}{\pi^{2}q^{2}})^{n-2}=\]\vspace*{-5mm}
\[=-\frac{3}{8\pi^{5}}\int\,\frac{d^{3}q}{q^{3}}\frac{1}{2\pi}\int\limits_{-\infty}^{\infty}du\times\]\vspace*{-5mm}
\begin{equation}
\times\sum_{n=2}^{\infty}\frac{(-1)^{n}}{n}[\hat{f}]^{n-1}[Q_{q}(u)]^{n}\left(\!\frac{\alpha\,r_{s}}{\pi^{2}q^{2}}\!\right)^{n-2}\!,
\end{equation}
where $\hat{F}$ is selected from the sum of four operators in Eq.
(14), which consist of four products of the operators of creation
and annihilation of particles, for instance:
$\hat{F}=\hat{a}_{\textbf{p}}^{+}\hat{a}_{\textbf{k}-\textbf{q}}^{+}\hat{a}_{\textbf{p}-\textbf{q}}\hat{a}_{\textbf{k}}$.
Then we obtain
\begin{equation}
\hat{f}=\hat{n}_{p}\hat{n}_{k+q}(1-\hat{n}_{p+q})(1-\hat{n}_{k}).
\end{equation}

In this section, we derive the equation of motion for the
mean value of the product
$\hat{b}_{-\textbf{p}}\hat{a}_{\textbf{p}}$ for a microscopic dipole,
which specifies of a medium polarization, which becomes macroscopic
due to the applied external field.

The average value of a certain physical magnitude $F$, which
corresponds to the operator $\hat{F}$ can be expressed through the spur of
a matrix, which is a certain statistic operator obeying the Heisenberg equation:
\[\langle\,\hat{F}\rangle={\rm Sp}(\hat{w}_{0}\hat{F})+\]\vspace*{-5mm}
\begin{equation}
+\frac{2\pi}{i}D(-\epsilon_{p_{1}+q}-\epsilon_{p_{2}-q}+\epsilon_{p_{1}}+\epsilon_{p_{2}}){\rm
Sp}([\hat{F},\hat{V}_{0}]\hat{w}_{0}),
\end{equation}
where $\hat{w}_{0}=\frac{e^{-\hat{H}_{0}/kT}}{{\rm
Sp}(e^{-\hat{H}_{0}/kT})}$, i.e., the density matrix $\hat{w}_{0}$
is assumed to be described by the Gibbs canonical distribution; in
the interaction representation, the time dependences of a wave
function and any certain operator can be expressed through the
Hamiltonian of a system of noninteracting particles: $\hat{V}_{0}=$
$=e^{i\hat{H}_{0}t/\hbar}\hat{V}e^{-i\hat{H}_{0}t/\hbar}$.

The Heisenberg equation for the electron-hole gas
takes the form
\[
\frac{d\hat{p}_{\textbf{p}}^{\nu_{e}\nu_{h}}}{dt}=-i\omega_{\textbf{p}}^{\nu_{e}\nu_{h}}\hat{p}_{\textbf{p}}^{\nu_{e}\nu_{h}}-i\Omega_{\textbf{p}}^{\nu_{e}\nu_{h}}(-1+\hat{n}_{\textbf{p}}^{\nu_{e}}+\hat{n}_{\textbf{p}}^{\nu_{h}})+\]\vspace*{-5mm}
\[+\frac{i}{\hbar}(\sum_{\textbf{q},\textbf{k}}V_{q}^{\nu_{e}\nu_{e}\nu_{e}\nu_{e}}\langle\,\hat{a}_{\textbf{k}+\textbf{q}}^{+}\hat{a}_{\textbf{p}+\textbf{q}}\hat{b}_{-\textbf{p}}\hat{a}_{\textbf{k}}\rangle +\]\vspace*{-5mm}
\[+V_{q}^{\nu_{h}\nu_{h}\nu_{h}\nu_{h}}\langle\,\hat{b}_{\textbf{k}+\textbf{q}}^{+}\hat{b}_{-\textbf{p}+\textbf{q}}\hat{b}_{\textbf{k}}\hat{a}_{\textbf{p}}\rangle-\]\vspace*{-5mm}
\[-\sum_{\textbf{q},\textbf{k}}V_{q}^{\nu_{e}\nu_{h}\nu_{h}\nu_{e}}(\langle\,\hat{a}_{\textbf{k}+\textbf{q}}^{+}\hat{a}_{\textbf{p}}\hat{b}_{-\textbf{p}+\textbf{q}}\hat{a}_{\textbf{k}}\rangle +\]\vspace*{-5mm}
\begin{equation}
+\langle\,\hat{b}_{\textbf{k}+\textbf{q}}^{+}\hat{b}_{-\textbf{p}}\hat{b}_{\textbf{k}}\hat{a}_{\textbf{p}+\textbf{q}}\rangle-\langle\,\hat{b}_{-\textbf{p}+\textbf{q}}\hat{a}_{\textbf{p}-\textbf{q}}\rangle\,\delta_{\textbf{q},\textbf{k}})),
\end{equation}
where
$\omega_{\textbf{p}}^{\nu_{e}\nu_{h}}=\frac{1}{\hbar}(\epsilon_{g0}+\epsilon_{e,\textbf{p}}^{\nu_{e}}+\epsilon_{h,\textbf{p}}^{\nu_{h}})$,
$\Omega_{\textbf{p}}^{\nu_{e}\nu_{h}}=$\linebreak
$=\frac{1}{\hbar}\mu_{\textbf{p}}^{\nu_{e}\nu_{h}}Ee^{-i\omega\,t}$.
Using the operator algebra and the density matrix formalism, we have
\[\frac{d\hat{p}_{\textbf{p}}^{\nu_{e}\nu_{h}}}{dt}=-i\omega_{\textbf{p}}^{\nu_{e}\nu_{h}}\hat{p}_{\textbf{p}}^{\nu_{e}\nu_{h}}-i\Omega_{\textbf{p}}^{\nu_{e}\nu_{h}}(-1+\hat{n}_{\textbf{p}}^{\nu_{e}}+\hat{n}_{\textbf{p}}^{\nu_{h}})-\]\vspace*{-5mm}
\[-\frac{i}{\hbar}\sum_{\textbf{q}}V_{q}^{\nu_{e}\nu_{h}\nu_{h}\nu_{e}}\hat{p}_{\textbf{p}+\textbf{q}}^{\nu_{e}\nu_{h}}(-1+\hat{n}_{\textbf{p}}^{\nu_{e}}+\hat{n}_{\textbf{p}}^{\nu_{h}})-\]\vspace*{-5mm}
\[-\frac{i}{\hbar}\sum_{\textbf{q}}W_{q}^{\nu_{e}\nu_{h}\nu_{h}\nu_{e}}\hat{p}_{\textbf{p}+\textbf{q}}^{\nu_{e}\nu_{h}}(\Xi_{\textbf{p},\textbf{q}}^{\nu_{e}}+\Xi_{\textbf{p},\textbf{q}}^{\nu_{h}})+\]\vspace*{-5mm}
\[+\frac{1}{\hbar}\sum_{\scriptsize\begin{array}{c}
\alpha=e,h\\
\beta=e,h\\
\alpha\neq\,\beta
\end{array}}\sum_{\nu_{\alpha},\nu_{\beta}}\sum_{\textbf{k},\textbf{q}}W_{q}^{\nu_{\alpha}\nu_{\beta}\nu_{\beta}\nu_{\alpha}}W_{|\textbf{p}+\textbf{q}-\textbf{k}|}^{\nu_{\alpha}\nu_{\beta}\nu_{\beta}\nu_{\alpha}}\times\]\vspace*{-5mm}
\[\times\,D(\epsilon_{\textbf{p}}^{\nu_{\beta}}+\epsilon_{\textbf{k}}^{\nu_{\alpha}}-\epsilon_{\textbf{k}-\textbf{q}}^{\nu_{\beta}}-\epsilon_{\textbf{p}+\textbf{q}}^{\nu_{\alpha}})\times\]\vspace*{-7mm}
\begin{equation}
\times(\hat{n}_{\textbf{p}}^{\nu_{\beta}}(1-\hat{n}_{\textbf{k}-\textbf{q}}^{\nu_{\beta}})\hat{n}_{\textbf{k}}^{\nu_{\alpha}}+(1-\hat{n}_{\textbf{p}}^{\nu_{\beta}})\hat{n}_{\textbf{k}-\textbf{q}}^{\nu_{\beta}}(1-\hat{n}_{\textbf{k}}^{\nu_{\alpha}}))\hat{p}_{\textbf{p}+\textbf{q}}^{\nu_{e}\nu_{h}}.
\end{equation}
Equation (26) describes the oscillation of the polarization at the
transition frequency and the processes of stimulated emission or
absorption. As the population functions, we choose the Fermi
distribution functions. The transition frequency
$\omega_{\textbf{p}}^{\nu_{e}\nu_{h}}$ is derived as follows:
\[\omega_{\textbf{p}}^{\nu_{e}\nu_{h}}=\frac{1}{\hbar}(\epsilon_{g0}+\epsilon_{e,\textbf{p}}^{\nu_{e}}+\epsilon_{h,\textbf{p}}^{\nu_{h}}+\]
\[+\sum_{\alpha=e,h}\sum_{\nu_{\alpha}}\sum_{\textbf{q}}(V_{q}^{\nu_{\alpha}\nu_{\alpha}\nu_{\alpha}\nu_{\alpha}}(-\hat{n}_{\textbf{p}+\textbf{q}}^{\nu_{\alpha}})+\]\vspace*{-5mm}
\[+W_{q}^{\nu_{\alpha}\nu_{\alpha}\nu_{\alpha}\nu_{\alpha}}(-\hat{\Xi}_{\textbf{p}+\textbf{q,q}}^{\nu_{\alpha}}))-\]\vspace*{-5mm}
\[-i\sum_{\scriptsize\begin{array}{c}
\alpha=e,h\\
\beta=e,h\\
\alpha\neq\,\beta
\end{array}}\sum_{\nu_{\alpha},\nu_{\beta}}\sum_{\textbf{k},\textbf{q}}(W_{q}^{\nu_{\alpha}\nu_{\beta}\nu_{\beta}\nu_{\alpha}})^{2}\times\]\vspace*{-5mm}
\[\times\,D(-\epsilon_{\textbf{p}+\textbf{q}}^{\nu_{\alpha}}-\epsilon_{\textbf{k}-\textbf{q}}^{\nu_{\beta}}+\epsilon_{\textbf{k}}^{\nu_{\beta}}+\epsilon_{\textbf{p}}^{\nu_{\alpha}})\times\]\vspace*{-5mm}
\begin{equation}
\times(\hat{n}_{\textbf{k}-\textbf{q}}^{\nu_{\beta}}(1-\hat{n}_{\textbf{k}}^{\nu_{\beta}})\hat{n}_{\textbf{p}+\textbf{q}}^{\nu_{\alpha}}+(1-\hat{n}_{\textbf{k}-\textbf{q}}^{\nu_{\beta}})\hat{n}_{\textbf{k}}^{\nu_{\beta}}(1-\hat{n}_{\textbf{p}+\textbf{q}}^{\nu_{\alpha}})).
\end{equation}
The functions $\hat{\Xi}_{\textbf{p}+\textbf{q,q}}^{\nu_{e}}$ and
$\hat{\Xi}_{\textbf{p,q}}^{\nu_{e}}$ are defined as
\[\hat{\Xi}_{\textbf{p}+\textbf{q,q}}^{\nu_{e}}=i\sum_{\textbf{k}}[W_{q}^{\nu_{e}\nu_{e}\nu_{e}\nu_{e}}-W_{|\textbf{k}-\textbf{q}-\textbf{p}|}^{\nu_{e}\nu_{e}\nu_{e}\nu_{e}}]\times\]\vspace*{-5mm}
\[\times\,D(-\epsilon_{\textbf{p}+\textbf{q}}^{\nu_{e}}-\epsilon_{\textbf{k}-\textbf{q}}^{\nu_{e}}+\epsilon_{\textbf{k}}^{\nu_{e}}+\epsilon_{\textbf{p}}^{\nu_{e}})\times\]\vspace*{-5mm}
\begin{equation}
\times(\hat{n}_{\textbf{k}-\textbf{q}}^{\nu_{e}}(1-\hat{n}_{\textbf{k}}^{\nu_{e}})\hat{n}_{\textbf{p}+\textbf{q}}^{\nu_{e}}+(1-\hat{n}_{\textbf{k}-\textbf{q}}^{\nu_{e}})\hat{n}_{\textbf{k}}^{\nu_{e}}(1-\hat{n}_{\textbf{p}+\textbf{q}}^{\nu_{e}}))\,
,
\end{equation}\vspace*{-7mm}
\[\hat{\Xi}_{\textbf{p,q}}^{\nu_{e}}=i\sum_{\textbf{k}}[W_{q}^{\nu_{e}\nu_{e}\nu_{e}\nu_{e}}-W_{|\textbf{k+q-p}|}^{\nu_{e}\nu_{e}\nu_{e}\nu_{e}}]\times\]\vspace*{-5mm}
\[\times\,D(-\epsilon_{\textbf{p}+\textbf{q}}^{\nu_{e}}-\epsilon_{\textbf{k}-\textbf{q}}^{\nu_{e}}+\epsilon_{\textbf{k}}^{\nu_{e}}+\epsilon_{\textbf{p}}^{\nu_{e}})\times\]
\begin{equation}
\times((1-\hat{n}_{\textbf{k}}^{\nu_{e}})\hat{n}_{\textbf{k}-\textbf{q}}^{\nu_{e}}(1-\hat{n}_{\textbf{p}}^{\nu_{e}})+\hat{n}_{\textbf{k}}^{\nu_{e}}(1-\hat{n}_{\textbf{k}-\textbf{q}}^{\nu_{e}})\hat{n}_{\textbf{p}}^{\nu_{e}}).
\end{equation}
We have replaced the bare Coulomb potential energy with the screened
one:
\begin{equation}
V_{q}(1-VM+(VM)^{2}-(VM)^{3}+...),
\end{equation}
where
\begin{equation}
M=\sum_{\textbf{k}}\frac{n(\epsilon_{\textbf{k}+\textbf{q}})-n(\epsilon_{\textbf{k}})}{\epsilon_{\textbf{k}+\textbf{q}}-\epsilon_{\textbf{k}}}.
\end{equation}
The coefficient of the sum in the second term of series (30) is
\[
N\frac{m}{2\hbar^{2}}\left(\!\frac{4\pi\,e^{2}}{\Omega}\!\right)^{\!2}\frac{\Omega}{(2\pi)^{3}}\frac{1}{k_{\rm
F}^{3}}2=\]\vspace*{-5mm}
\begin{equation}
=N\frac{me^{4}}{2\hbar^{2}}\left(\!\frac{4\pi}{\Omega}\!\right)^{\!2}\frac{\Omega}{(2\pi)^{3}}\frac{\Omega}{3\pi^{2}N}2=\frac{me^{4}}{2\hbar^{2}}\frac{4}{3\pi^{3}},
\end{equation}
In the third term of the series, the coefficient is
\[
\frac{m^{2}}{2\hbar^{4}}\left(\!\frac{4\pi\,e^{2}}{\Omega}\!\right)^{\!3}\left(\!\frac{\Omega}{(2\pi)^{3}}\!\right)^{\!2}\frac{1}{k_{\rm
F}^{3}}\frac{1}{k_{\rm
F}}2=\frac{me^{4}}{2\hbar^{2}}\frac{4}{3\pi^{3}}\frac{\alpha\,r_{s}}{2\pi^{2}}.\]\vspace*{-5mm}
\begin{equation}
\alpha\,r_{s}=\frac{me^{2}}{\hbar^{2}}\frac{1}{k_{\rm F}}.
\end{equation}
Then the series can be rewritten as a infinitely decreasing
geometric progression
\[\frac{1}{q^{2}}-\frac{4}{3\pi^{3}}\int\,d^{3}k\frac{1}{q^{4}}\frac{n(\epsilon_{\textbf{k}+\textbf{q}})-n(\epsilon_{\textbf{k}})}{(\textbf{k}+\textbf{q})^{2}-k^{2}}+\]\vspace*{-5mm}
\[+\frac{4}{3\pi^{3}}\frac{\alpha\,r_{s}}{2\pi^{2}}\int\int\,d^{3}k_{1}d^{3}k_{2}\frac{1}{q^{6}}\frac{n(\epsilon_{\textbf{k}_{1}+\textbf{q}})-n(\epsilon_{\textbf{k}_{1}})}{(\textbf{k}_{1}+\textbf{q})^{2}-k_{1}^{2}}\times\]\vspace*{-5mm}
\begin{equation}
\times\frac{n(\epsilon_{\textbf{k}_{2}+\textbf{q}})-n(\epsilon_{\textbf{k}_{2}})}{(\textbf{k}_{2}+\textbf{q})^{2}-k_{2}^{2}}-...\,
.
\end{equation}
By summing all terms of the series, we obtain
\begin{equation}
W_{q}^{\nu_{\alpha}\nu_{\beta}\nu_{\beta}\nu_{\alpha}}=\frac{V_{q}^{\nu_{\alpha}\nu_{\beta}\nu_{\beta}\nu_{\alpha}}}{\varepsilon_{q}(N)}.
\end{equation}
For the dielectric function, we use the static Lindhard
formula:\vspace*{-3mm}
\begin{equation}
\varepsilon_{q}(N)=1-\sum_{\rho=e,h}\sum_{\nu_{\rho}}\sum_{\textbf{p}}V_{q}^{\nu_{\rho}\nu_{\rho}\nu_{\rho}\nu_{\rho}}\frac{\hat{n}_{\textbf{p}+\textbf{q}}^{\nu_{\rho}}-\hat{n}_{\textbf{p}}^{\nu_{\rho}}}{\epsilon_{\textbf{p}+\textbf{q}}^{\nu_{\rho}}-\epsilon_{\textbf{p}}^{\nu_{\rho}}}.
\end{equation}
Since the cyclic diagrams are the basic type of diagrams in the
scattering processes at a high density of the electron-hole gas, the
diagram method is equivalent of the self-consistency method, as well
as the random phase approximation.

The answer how to derive the integro-differential equation (26) for
a microscopic dipole is given by the scheme
\begin{equation}
\omega_{\textbf{p}}^{\nu_{e}\nu_{h}}:\,V_{q}^{\nu_{\alpha}\nu_{\alpha}\nu_{\alpha}\nu_{\alpha}}\rightarrow\,W_{q}^{\nu_{\alpha}\nu_{\alpha}\nu_{\alpha}\nu_{\alpha}}\,,n_{\textbf{p}+\textbf{q}}^{\nu_{\alpha}}\rightarrow\,\Xi_{\textbf{p+q,q}}^{\nu_{\alpha}},
\end{equation}
plus the expression, whose graphic representation reminds a binary
blister,
\begin{equation}
\sum_{\textbf{p}}\frac{d\hat{p}_{\textbf{p}}^{\nu_{e}\nu_{h}}}{dt}:\,V_{q}^{\nu_{e}\nu_{h}\nu_{h}\nu_{e}}\rightarrow\,W_{q}^{\nu_{e}\nu_{h}\nu_{h}\nu_{e}}\,,n_{\textbf{p}}^{\nu_{\alpha}}\rightarrow\,\Xi_{\textbf{p,q}}^{\nu_{\alpha}},
\end{equation}\vspace*{-5mm}

\noindent plus the expression corresponding to the plot in the form
of an oyster.

The sum over momenta in the polarization equation, which includes
the carrier-carrier correlations of higher orders than Hartree--Fock
ones, can be found if the self-energy in the equation is added by
the term, which is present in the equation in the Hartree--Fock
approximation, by replacing the Coulomb potential energy with the
screened one and the Fermi distribution functions with the
$\Xi_{\textbf{p+q,q}}^{\nu_{\alpha}}$ functions, plus the
expression, whose schematic representation is in the form of a
binary blister. The integro-differential equation should be added by
the term which is present in the equation in the Hartree--Fock
approximation, by replacing the Coulomb potential energy with the
screened one and the Fermi distribution functions with the
$\Xi_{\textbf{p,q}}^{\nu_{\alpha}}$ functions, plus the expression,
whose schematic representation is in the form of an oyster. We
consider the coupled closed diagrams. The sum of all uncoupled
diagrams, which include $k$, closed loops which have
$m_{1},m_{2},...,m_{k}$ vertices, correspondingly, is the sum of all
closed diagrams of the $m$-th order.

The polarization equation written in the different designations was
obtained in \cite{{Chow}} and is divided into diagonal and
nondiagonal terms with respect to $p_{\textbf{p}}^{\nu_{e}\nu_{h}}$
\[\frac{d\hat{p}_{\textbf{p}}^{\nu_{e}\nu_{h}}}{dt}=-i\omega_{\textbf{p}}^{\nu_{e}\nu_{h}}\hat{p}_{\textbf{p}}^{\nu_{e}\nu_{h}}-i\Omega_{\textbf{p}}^{\nu_{e}\nu_{h}}(-1+\hat{n}_{\textbf{p}}^{\nu_{e}}+\hat{n}_{\textbf{p}}^{\nu_{h}})+\]
\vspace*{-5mm}
\begin{equation}
+(\Gamma_{\textbf{p}}^{\nu_{e}}+\Gamma_{\textbf{p}}^{\nu_{h}})\hat{p}_{\textbf{p}}^{\nu_{e}\nu_{h}}+\sum_{\textbf{q}}(\Gamma_{\textbf{pq}}^{\nu_{e}}+\Gamma_{\textbf{pq}}^{\nu_{h}})\hat{p}_{\textbf{p}+\textbf{q}}^{\nu_{e}\nu_{h}}.
\end{equation}
The transition frequency $\omega_{\textbf{p}}^{\nu_{e}\nu_{h}}$  and
the Rabi frequency are derived as follows:
\[
\omega_{\textbf{p}}^{\nu_{e}\nu_{h}}=\frac{1}{\hbar}(\epsilon_{g0}+\epsilon_{e,\textbf{p}}^{\nu_{e}}+\epsilon_{h,\textbf{p}}^{\nu_{h}}-\]
\vspace*{-5mm}
\begin{equation}
-\sum_{\alpha=e,h}\sum_{\textbf{q}}V_{q}^{\nu_{\alpha}\nu_{\alpha}\nu_{\alpha}\nu_{\alpha}}n_{\textbf{p}+\textbf{q}}^{\nu_{\alpha}}),
\end{equation}\vspace*{-5mm}
\begin{equation}
\Omega_{\textbf{p}}^{\nu_{e}\nu_{h}}=\frac{1}{\hbar}(\mu_{\textbf{p}}^{\nu_{e}\nu_{h}}Ee^{-i\omega\,t}+\sum_{\textbf{q}}V_{q}^{\nu_{e}\nu_{h}\nu_{h}\nu_{e}})\hat{p}_{\textbf{p}+\textbf{q}}^{\nu_{e}\nu_{h}}.
\end{equation}
Carrier-carrier correlations which lead to the screening and the
dephasing are described by the expressions which include the
diagonal ($p_{\textbf{p}}^{\nu_{e}\nu_{h}}$ terms) and nondiagonal
($p_{\textbf{p}+\textbf{q}}^{\nu_{e}\nu_{h}}$ terms) contributions.
For the diagonal contribution,
\[
\Gamma_{\textbf{p}}^{\nu_{\alpha}}=-\frac{2\pi}{\hbar}\sum_{\beta=e,h}\sum_{\nu_{\beta}}\sum_{\textbf{k,q}}(|W_{q}^{\nu_{\alpha}\nu_{\beta}\nu_{\beta}\nu_{\alpha}}|^{2}-\]\vspace*{-5mm}
\[-\frac{1}{2}\delta_{\nu_{\alpha}\nu_{\beta}}W_{q}^{\nu_{\alpha}\nu_{\beta}\nu_{\beta}\nu_{\alpha}}W_{|\textbf{k}-\textbf{q}-\textbf{p}|}^{\nu_{\alpha}\nu_{\beta}\nu_{\beta}\nu_{\alpha}})\times\]\vspace*{-5mm}
\[\times\,D(-\epsilon_{\textbf{p}+\textbf{q}}^{\nu_{\alpha}}-\epsilon_{\textbf{k}-\textbf{q}}^{\nu_{\beta}}+\epsilon_{\textbf{k}}^{\nu_{\beta}}+\epsilon_{\textbf{p}}^{\nu_{\alpha}})\times\]
\begin{equation}
\times(\hat{n}_{\textbf{k}-\textbf{q}}^{\nu_{\beta}}(1-\hat{n}_{\textbf{k}}^{\nu_{\beta}})\hat{n}_{\textbf{p}+\textbf{q}}^{\nu_{\alpha}}+(1-\hat{n}_{\textbf{k}-\textbf{q}}^{\nu_{\beta}})\hat{n}_{\textbf{k}}^{\nu_{\beta}}(1-\hat{n}_{\textbf{p}+\textbf{q}}^{\nu_{\alpha}})).
\end{equation}
There are also the nondiagonal contributions which couple the
polarizations for different wave vectors and are defined by the
expression
\[
\Gamma_{\textbf{qp}}^{\nu_{\alpha}}=-\frac{2\pi}{\hbar}\sum_{\scriptsize\begin{array}{c}
\beta=e,h\\
\beta'=e,h\\
\beta\neq\,\alpha
\end{array}}\sum_{\nu_{\beta},\nu_{\beta'}}\sum_{\textbf{k}}(|W_{q}^{\nu_{\alpha}\nu_{\beta}\nu_{\beta}\nu_{\alpha}}|^{2}-\]\vspace*{-5mm}
\[-W_{q}^{\nu_{\alpha}\nu_{\beta}\nu_{\beta}\nu_{\alpha}}W_{|\textbf{p}+\textbf{q}-\textbf{k}|}^{\nu_{\alpha}\nu_{\beta'}\nu_{\beta'}\nu_{\alpha}}+\]\vspace*{-5mm}
\[+\frac{1}{2}\delta_{\nu_{\alpha}\nu_{\beta'}}W_{q}^{\nu_{\alpha}\nu_{\beta}\nu_{\beta}\nu_{\alpha}}W_{|\textbf{k}+\textbf{q}-\textbf{p}|}^{\nu_{\alpha}\nu_{\beta'}\nu_{\beta'}\nu_{\alpha}})\times\]\vspace*{-5mm}
\[\times\,D(-\epsilon_{\textbf{k}}^{\nu_{\alpha}}-\epsilon_{\textbf{p}}^{\nu_{\beta'}}+\epsilon_{\textbf{k}-\textbf{q}}^{\nu_{\beta'}}+\epsilon_{\textbf{p}+\textbf{q}}^{\nu_{\alpha}})\times\]\vspace*{-5mm}
\begin{equation}
\times(\hat{n}_{\textbf{p}}^{\nu_{\beta'}}(1-\hat{n}_{\textbf{k}-\textbf{q}}^{\nu_{\beta'}})\hat{n}_{\textbf{k}}^{\nu_{\alpha}}+(1-\hat{n}_{\textbf{p}}^{\nu_{\beta'}})\hat{n}_{\textbf{k}-\textbf{q}}^{\nu_{\beta'}}(1-\hat{n}_{\textbf{k}}^{\nu_{\alpha}})),
\end{equation}
We solve the system of differential equations and derive the system of algebraic
equations, i.e., the integral equation
\[
\chi_{\textbf{p}}^{\nu_{e}\nu_{h}}=\frac{i}{\hbar}\frac{(\hat{n}_{\textbf{p}}^{\nu_{e}}+\hat{n}_{\textbf{p}}^{\nu_{h}}-1)}{i(\omega_{\textbf{p}}^{\nu_{e}\nu_{h}}-\omega)+\Gamma_{\textbf{p}}^{\nu_{e}}+\Gamma_{\textbf{p}}^{\nu_{h}}}(\mu_{\textbf{p}}^{\nu_{e}\nu_{h}}-\]\vspace*{-5mm}
\begin{equation}
-\sum_{\textbf{q}}V_{\scriptsize\left\{\begin{array}{c}
|-\textbf{p}|\\
|-\textbf{p}-\textbf{q}|\\
\end{array}\right\}}^{\nu_{e}\nu_{h}\nu_{h}\nu_{e}}\chi_{\textbf{p}+\textbf{q}}^{\nu_{e}\nu_{h}}),
\end{equation}
in which $\omega_{\textbf{p}}^{\nu_{e}\nu_{h}}$ is the self-energy,
i.e., the renormalized width of the band gap. The
derived renormalization energy is the exchange energy. The sum
$\Gamma_{\textbf{p}}^{\nu_{e}\nu_{h}}$ defines a half-width of
the exciton resonance. The polarization is expressed through the
function $\chi_{\textbf{p}}^{\nu_{e}\nu_{h}}$ as follows:
\begin{equation}
p_{\textbf{p}}^{\nu_{e}\nu_{h}}=\chi_{\textbf{p}}^{\nu_{e}\nu_{h}}Ee^{-i\omega\,t}.
\end{equation}

The half-width of the gain spectra is calculated with the help of
formulas
\[
\Gamma_{\textbf{k}}^{\nu_{\alpha}}=\frac{1}{2\pi\,\hbar}\sum_{\beta=e,h}\sum_{\nu_{\beta}}\int\limits_{0}^{2\pi}d\varphi\frac{1}{2\pi}\int\limits_{0}^{2\pi}d\alpha\times\]\vspace*{-5mm}
\[\times\int\,qdq\frac{1}{\displaystyle{\partial(\varepsilon_{\textbf{k}+\textbf{q}}^{\nu_{\beta}}-\varepsilon_{\textbf{k}+\textbf{q}}^{\nu_{\alpha}})}/{\partial|\textbf{k}+\textbf{q}|}}\times\]\vspace*{-5mm}
\[\times\,Q\Biggl(\int\,dz\int\,dz'\chi_{\nu_{\alpha}}(z)\chi_{\nu_{\beta}}(z')\times\]
\[\times\,e^{-q|z-z'|}\chi_{\nu_{\beta}}(z')\chi_{\nu_{\alpha}}(z)\frac{2\pi}{q}\!\Biggr)^{\!2}\times\]\vspace*{-5mm}
\begin{equation}
\times(\hat{n}_{\textbf{k}+\textbf{q}}^{\nu_{\alpha}}(1-\hat{n}_{\textbf{k}+\textbf{q}}^{\nu_{\beta}})\hat{n}_{\textbf{k}}^{\nu_{\beta}}+(1-\hat{n}_{\textbf{k}+\textbf{q}}^{\nu_{\alpha}})\hat{n}_{\textbf{k}+\textbf{q}}^{\nu_{\beta}}(1-\hat{n}_{\textbf{k}}^{\nu_{\beta}})),
\end{equation}\vspace*{-5mm}
\[
\Gamma_{\textbf{k}}^{\nu_{\alpha}}=\frac{1}{2\pi\,\hbar}\sum_{\beta=e,h}\sum_{\nu_{\beta}}\int\limits_{0}^{2\pi}d\varphi\frac{1}{2\pi}\int\limits_{0}^{2\pi}d\alpha\times\]\vspace*{-5mm}
\[\times\int\,qdq\frac{1}{\frac{\partial(\varepsilon_{\textbf{k}+\textbf{q}}^{\nu_{\alpha}})}{\partial|\textbf{k}+\textbf{q}|}}Q \Biggl(\!\Biggl(\!\int\,dz\int\,dz'\chi_{\nu_{\alpha}}(z)\chi_{\nu_{\beta}}(z')\times\]\vspace*{-5mm}
\[\times\,e^{-q|z-z'|}\chi_{\nu_{\beta}}(z')\chi_{\nu_{\alpha}}(z)\frac{2\pi}{q}\!\Biggr)^{\!2}-\]\vspace*{-5mm}
\[-\frac{1}{2}\delta_{\nu_{\alpha}\nu_{\beta}}\Biggl(\int\,dz\int\,dz'\chi_{\nu_{\alpha}}(z)\chi_{\nu_{\beta}}(z')\times\]\vspace*{-5mm}
\[\times\,e^{-q|z-z'|}\chi_{\nu_{\beta}}(z')\chi_{\nu_{\alpha}}(z)\frac{2\pi}{q}\times\]\vspace*{-5mm}
\[\times\int dz\int dz'\chi_{\nu_{\alpha}}(z)\chi_{\nu_{\beta}}(z')e^{-k\sqrt{2-2\cos(\alpha)}|z-z'|}\times\]\vspace*{-5mm}
\[\times\chi_{\nu_{\beta}}(z')\chi_{\nu_{\alpha}}(z)\frac{2\pi}{k\sqrt{2-2\cos(\alpha)}}\!\Biggr)\!\Biggr)\times\]\vspace*{-5mm}
\[\times(\hat{n}_{\textbf{k}+\textbf{q}}^{\nu_{\alpha}}(1-\hat{n}_{\textbf{k}+\textbf{q}}^{\nu_{\beta}})\hat{n}_{\textbf{k}}^{\nu_{\beta}}+\]\vspace*{-5mm}
\begin{equation}
+(1-\hat{n}_{\textbf{k}+\textbf{q}}^{\nu_{\alpha}})\hat{n}_{\textbf{k}+\textbf{q}}^{\nu_{\beta}}(1-\hat{n}_{\textbf{k}}^{\nu_{\beta}}))\delta_{\nu_{\alpha}\nu_{\beta}},
\end{equation}
where $Q=|\textbf{k}+\textbf{q}|$, $\varphi$ is the angle between
the vectors $\textbf{k}$ and $\textbf{q}$. In calculations of the
broadening caused by carrier-carrier correlations, one can see that, in
the schematic representation, their expressions in the form of
diagrams include two diagrams in the form of an oyster and four
expressions diagrams in the
form of a binary blister.

The polarization equation for the wurtzite quantum well in
the Hartree--Fock approximation with regard for the wave functions for an
electron and a hole written in the form \cite{{Lokot},{Lokot1}}, where
the coefficients of the expansion of the wave function of a hole
in the basis of wave functions (known as spherical harmonics) with the
orbital angular momentum $l=1$ and the eigenvalue $m_{l}$ depend on the wave vector, its $z$
component, can looked for as
follows:
\begin{equation}
\begin{array}{c}
\frac{d\hat{p}_{\textbf{p}}^{\nu_{e}\nu_{h}}}{dt}=-i\omega_{\textbf{p}}^{\nu_{e}\nu_{h}}\hat{p}_{\textbf{p}}^{\nu_{e}\nu_{h}}-i\Omega_{\textbf{p}}^{\nu_{e}\nu_{h}}(-1+\hat{n}_{\textbf{p}}^{\nu_{e}}+\hat{n}_{\textbf{p}}^{\nu_{h}}).\\
\end{array}
\end{equation}
The transition frequency $\omega_{\textbf{p}}^{\nu_{e}\nu_{h}}$ and
the Rabi frequency with regard for the wave function
\cite{{Lokot},{Lokot1}} are described as
\[
\omega_{\textbf{p}}^{\nu_{e}\nu_{h}}=\frac{1}{\hbar}(\epsilon_{g0}+\epsilon_{e,\textbf{p}}^{\nu_{e}}+\epsilon_{h,\textbf{p}}^{\nu_{h}}-\sum_{\textbf{q}}V_{q}^{\nu_{e}\nu_{e}\nu_{e}\nu_{e}}n_{\textbf{p}+\textbf{q}}^{\nu_{e}}-\]\vspace*{-5mm}
\begin{equation}
-\sum_{\textbf{q}}V_{\scriptsize\left\{\begin{array}{cc}
|-\textbf{p}+\textbf{q}| & |-\textbf{p}|\\
|-\textbf{p}+\textbf{q}| & |-\textbf{p}|\\
\end{array}\right\}}^{\nu_{h}\nu_{h}\nu_{h}\nu_{h}}n_{-\textbf{p}+\textbf{q}}^{\nu_{h}}),
\end{equation}\vspace*{-5mm}
\begin{equation}
\Omega_{\textbf{p}}^{\nu_{e}\nu_{h}}=\frac{1}{\hbar}(\mu_{\textbf{p}}^{\nu_{e}\nu_{h}}Ee^{-i\omega\,t}+\sum_{\textbf{q}}V_{\scriptsize\left\{\begin{array}{c}
|-\textbf{p}|\\
|-\textbf{p}-\textbf{q}|\\
\end{array}\right\}}^{\nu_{e}\nu_{h}\nu_{h}\nu_{e}})\hat{p}_{\textbf{p}+\textbf{q}}^{\nu_{e}\nu_{h}},\\
\end{equation}\vspace*{-7mm}

\noindent where\vspace*{-3mm}
\[
V_{\scriptsize\left\{\begin{array}{c}
|-\textbf{p}|\\
|-\textbf{p}-\textbf{q}|\\
\end{array}\right\}}^{\nu_{e}\nu_{h}\nu_{h}\nu_{e}}=\frac{1}{2}\frac{e^{2}}{\varepsilon}\frac{1}{2\pi}\int\limits_{0}^{2\pi}d\varphi\sum_{\alpha}g_{\alpha}\int dq\times\]\vspace*{-5mm}
\[\times\int\,dz_{\xi}\int\,dz_{\xi'}\chi_{n_{1}}(z_{\xi})\chi_{m_{1}}(z_{\xi'})\chi_{m_{2}}(z_{\xi'})\chi_{n_{2}}(z_{\xi})\times\]\vspace*{-5mm}
\[
\times\,e^{-q|z_{\xi}-z_{\xi'}|}C_{p}^{j}[n_{1},1]V_{p}^{j}[m_{1},1]C_{Q_{1}}^{i}[n_{2},1]V_{Q_{1}}^{i}[m_{2},1],\]\vspace*{-7mm}
\[n_{1}=m_{1}=n_{2}=m_{2}=1,\]\vspace*{-7mm}
\begin{equation}
\textbf{Q}_{1}=\textbf{q}+\textbf{p},
\end{equation}\vspace*{-5mm}
\[
\sum_{\alpha,\textbf{q}}g_{\alpha}V_{\scriptsize\left\{\!\begin{array}{cc}
|-\textbf{p}+\textbf{q}| & |-\textbf{p}|\\
|-\textbf{p}+\textbf{q}| & |-\textbf{p}|\\
\end{array}\!\right\}}^{\nu_{h}\nu_{h}\nu_{h}\nu_{h}}n_{-\textbf{p}+\textbf{q}}^{\nu_{h}}\!=\!\frac{1}{2}\frac{e^{2}}{\varepsilon}\frac{1}{2\pi}\!\int\limits_{0}^{2\pi}\!\!d\varphi\!\sum_{\alpha}\!g_{\alpha}\times\]\vspace*{-5mm}
\[\times\int\! dq\int\! dz_{\xi}\int\! dz_{\xi'}\chi_{n_{1}}(z_{\xi})\chi_{m_{1}}(z_{\xi'})\chi_{m_{2}}(z_{\xi'})\chi_{n_{2}}(z_{\xi})\times\]\vspace*{-5mm}
\[\times\,e^{-q|z_{\xi}-z_{\xi'}|}V_{Q_{2}}^{j}[n_{1},1]V_{p}^{i}[m_{1},1]\times\]\vspace*{-7mm}
\[\times V_{Q_{2}}^{i}[n_{2},1]V_{p}^{j}[m_{2},1]n_{\alpha,Q_{2}},\]\vspace*{-7mm}
\begin{equation}
\textbf{Q}_{2}=\textbf{q}-\textbf{p},
\end{equation}
where $\chi_{n_{1}}(z_{\xi})$ is the envelope of the wave
functions of the quantum well, $V_{p}^{i}[m_{1},1]$ and
$C_{p}^{j}[n_{1},1]$  are coefficients of the expansion of the wave
functions of a hole and electron at the envelope part, $\varphi$ is
the angle between the vectors $\textbf{p}$ and $\textbf{q}$, and
$g_{\alpha}$ is a degeneracy order of a level.

Numerically solving this integro-differential equation, we can
obtain the absorption coefficient of a plane wave in the medium from
the Maxwell equations:
\begin{equation}
\alpha(\omega)=\frac{\omega}{\varepsilon_{0}\,ncE}{\rm Im}\,P,
\end{equation}
where $\varepsilon_{0}$ and $c$  are the permittivity and the
speed of light, respectively, in vacuum, $n$  is a background
refractive index of the quantum well material,
\begin{equation}
P=\frac{2}{A}\sum_{\nu_{e},\nu_{h},\textbf{p}}(\mu_{\textbf{p}}^{\nu_{e}\nu_{h}})^{\star}p_{\textbf{p}}^{\nu_{e}\nu_{h}}e^{i\omega\,t}.
\end{equation}

\begin{figure}
\includegraphics[width=\column]{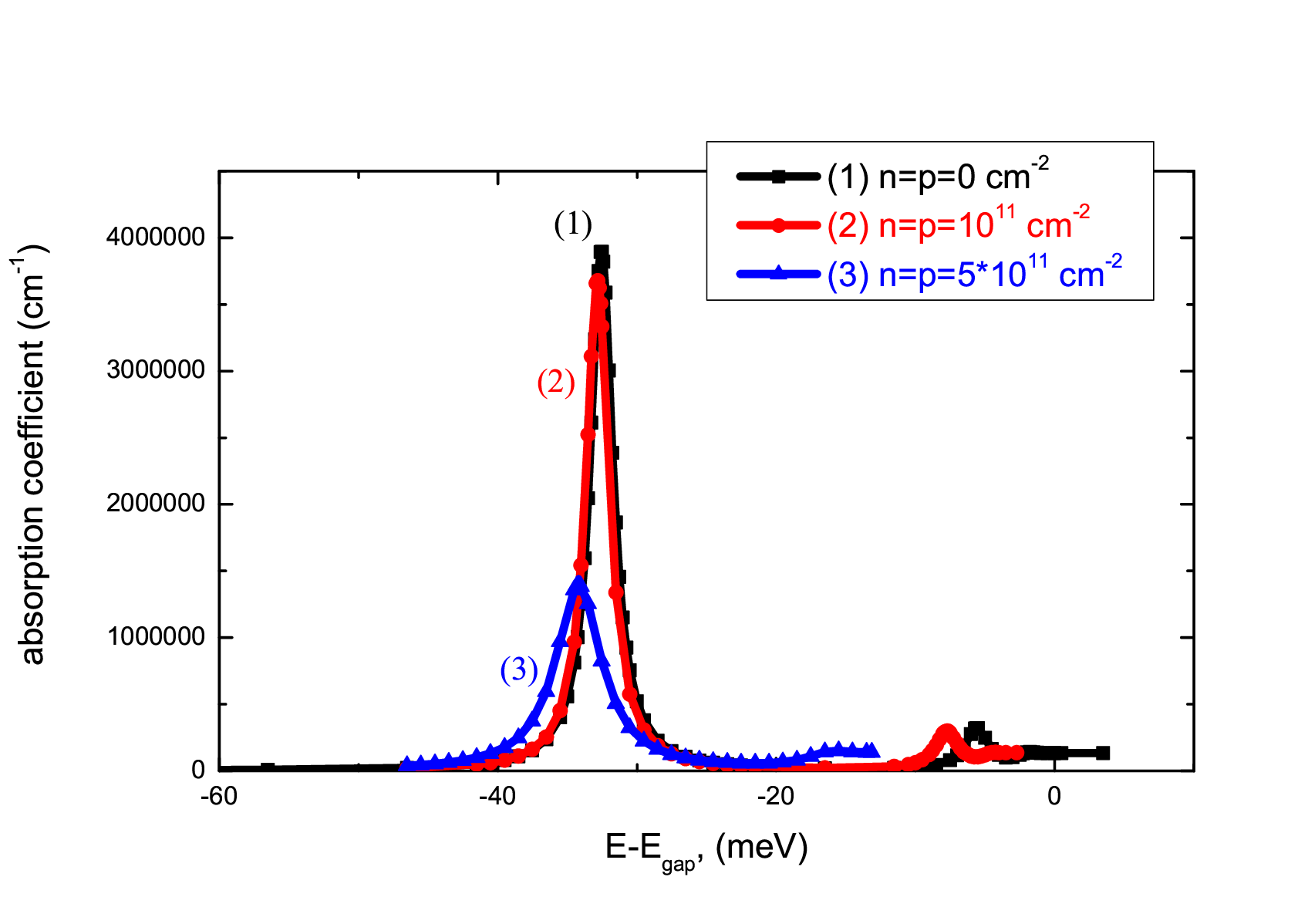}
\vskip-2mm\caption{Calculated Hartree--Fock spectra for the quantum
well with a width of 2.6 nm  }\vskip3mm
\end{figure}
\begin{figure}
\includegraphics[width=\column]{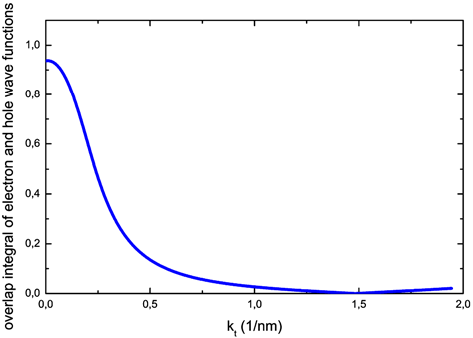}
\vskip-2mm\caption{Overlap integral of the electron and hole wave
functions  }\vskip3mm
\end{figure}
\begin{figure}
\includegraphics[width=\column]{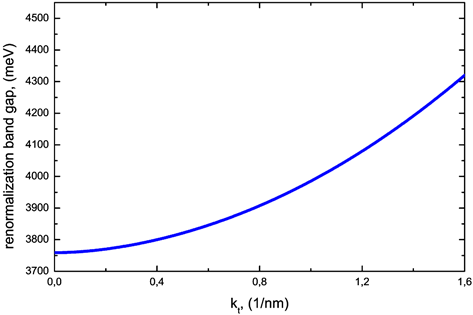}
\vskip-3mm\caption{Dispersion of the renormalization band gap for
the quantum well with a width of 2 nm at the concentration of
carriers $5\times10^{11}$ cm$^{-2}$  }
\end{figure}

\section{Results and Their Discussions}

Numerically solving the microscopic polarization equations for the
quantum well with a parabolic band, one can see that, with
increasing the electron-hole gas density, the optical gain develops
in the spectral region of the original exciton resonance. With
increasing the free-carrier density, the ionization continuum shifts
rapidly to longer wavelengths, while the 1$s$-exciton absorption
line stays almost constant, due to the high degree of compensation
between the weakening of the electron-hole binding energy and the
band-gap reduction. Physically, this indicates the charge neutrality
of an exciton~\cite{{Haug}}. The exciton absorption spectrum for the
quantum well with a parabolic law of dispersion is presented in
Fig.~1. All calculations are carried at a temperature
of~300~K.

The overlap integral of the electron and hole wave functions is
presented in Fig. 2.

Unlike will be develop the process of shifting of the absorption
edge at a constant exciton energy with increasing
the concentrations for the wurtzite quantum well. Solving the polarization
equation in the Hartree--Fock approximation, one can find a red shift of the
exciton resonance with increasing the concentration in the wurtzite
quantum well. The calculated Hartree--Fock spectrum for the wurtzite
quantum well with a width of 2 nm is presented in Fig. 4.

Such a shift can be explained by the difference between the overlap
integrals of the electron and hole wave functions in the wurtzite
quantum well and the quantum well with a parabolic band. The overlap
integral of the electron and hole wave functions at nonzero wave
vectors in the wurtzite quantum well has a smaller value than the
overlap integral in the quantum well with a parabolic band. Due to
this cause, the Coulomb renormalization of the electric dipole
moment in (50) in the wurtzite quantum well is found to be smaller
than that in the quantum well with a parabolic band and cannot
compensate the Coulomb renormalization of the self-energy in (49),
where it has the minus sign. This yields a shift of the exciton
resonance to the side of less energies. Since the shift of the
exciton resonance is a very rare effect, the examples of exceptions
are always in\-teresting.\looseness=1

\begin{figure}
\includegraphics[width=\column]{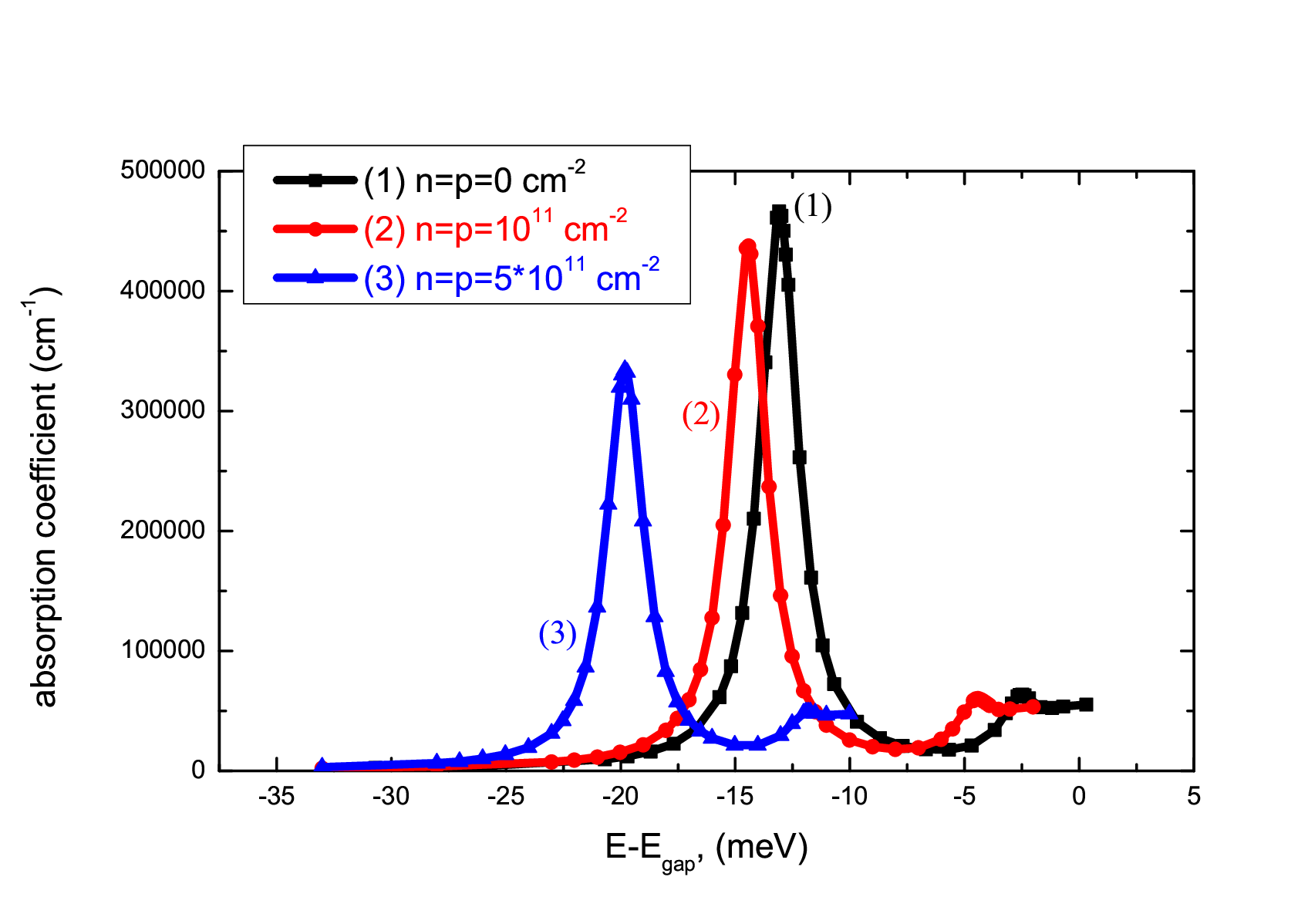}
\vskip-3mm\caption{Calculated Hartree--Fock spectra for the quantum
well with a width of 2 nm  }\vskip3mm
\end{figure}
\begin{figure}
\includegraphics[width=\column]{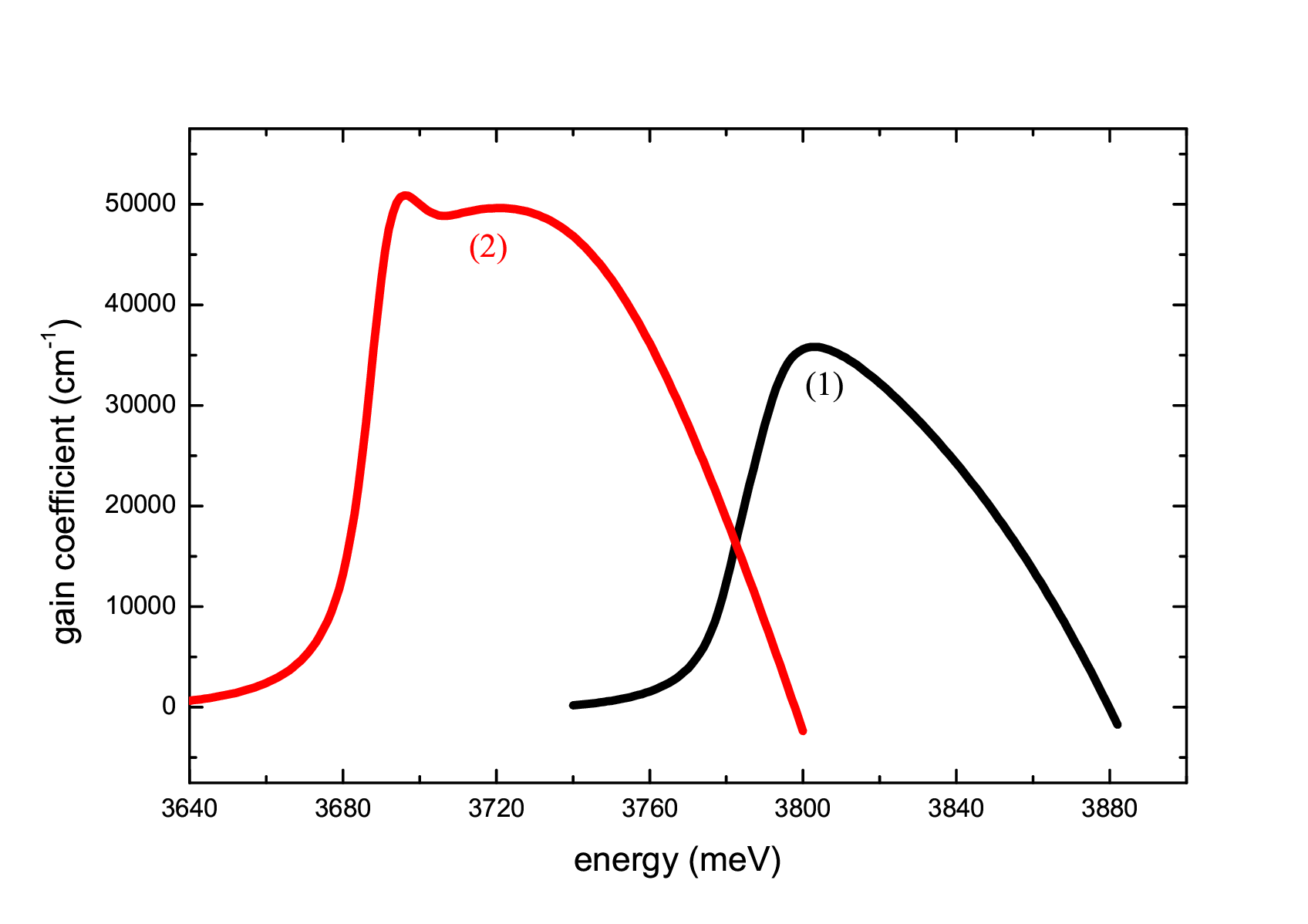}
\vskip-3mm\caption{Hartree gain spectrum (1) and Hartree--Fock gain
spectrum (2) at the concentration of carriers $n=p=9\times10^{12}$
cm$^{-2}$ for the quantum well with a width of 2 nm at the
temperature 300~K  }
\end{figure}

The dispersion of the renormalization band gap for the quantum well
with a width of 2 nm at the concentration of carrier
$5\times10^{11}$ cm$^{-2}$ is presented in Fig.~3. The energy of the
exciton resonance is calculated, and it is found that, for the
concentration of carriers $5\times10^{11}$ cm$^{-2}$, the exciton
energy is equal to 3749.5 meV.

In general, the existence of the resonance and the Sommerfeld enhancement of
a continuous optical spectrum is a reflection of the
renormalization of the electric dipole interaction energy and is
a cause of increasing the optical absorption in comparing with
the optical spectrum of free carriers. This increase of the
absorption is the example of a more general phenomenon of Coulomb
enhancement and can be explained as follows. Due to the Coulomb attraction, an
electron and a hole have a larger tendency to be located closer to
each other, as compared with the case of noninteracting particles.
This increase of the interaction duration leads to an increase of the
optical transition probability and to the renormalization of the electric
dipole interaction energy.

The Hartree and Hartree--Fock gain spectra are presented in Fig. 5.

\begin{figure}
\includegraphics[width=7.8cm]{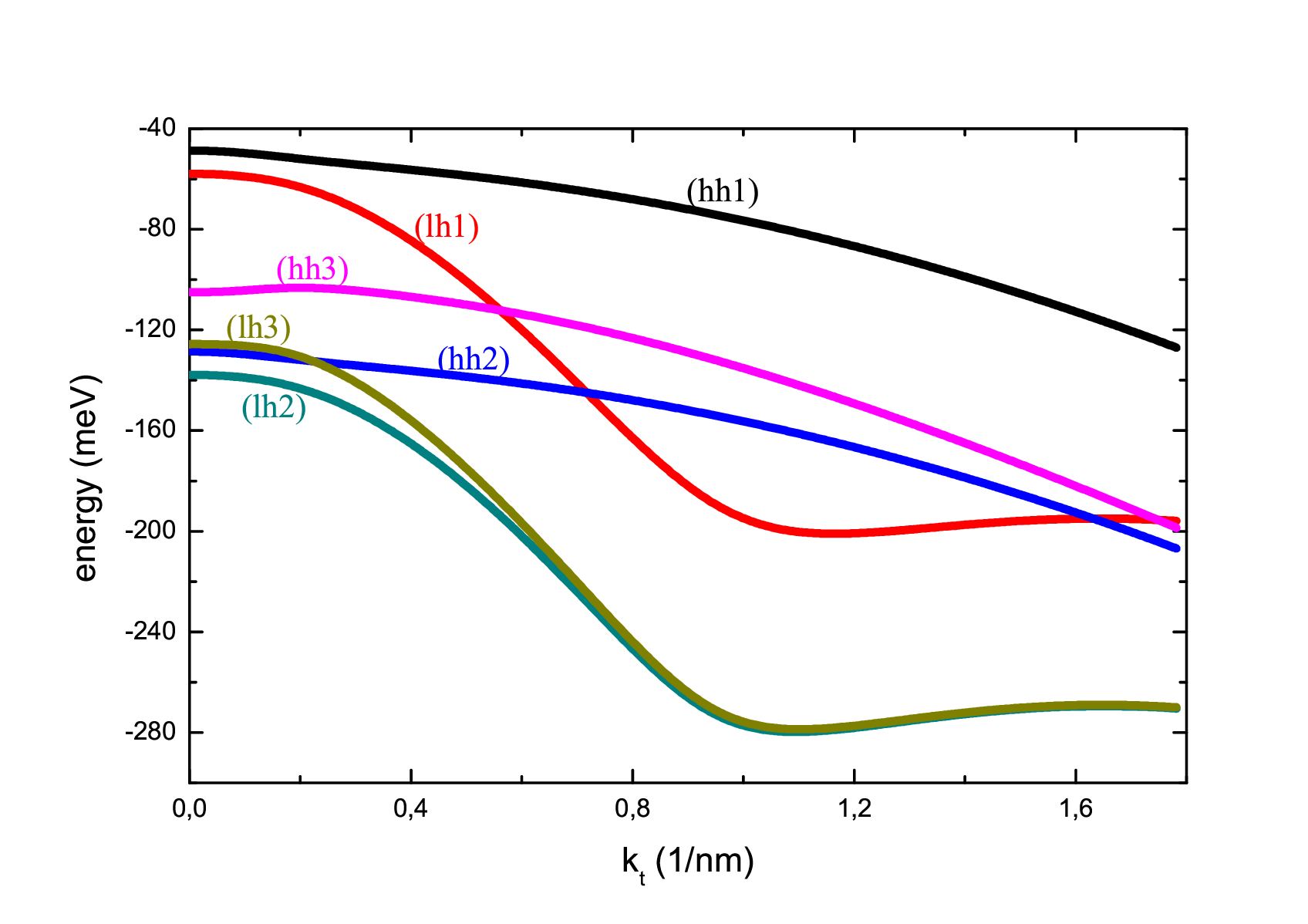}
\vskip-3mm\caption{Calculated energy spectra for heavy (hh1) and
light (lh1) holes for the free valence band, Hartree energy spectra
for heavy (hh2) and light (lh2) holes, and Hartree--Fock energy
spectra for the heavy (hh3) and light (lh3) holes for the quantum
well with a width of 2 nm at the concentration of carriers
$n=p=9\times10^{12}$ cm$^{-2}$ at the temperature 300~K  }\vskip3mm
\end{figure}
\begin{figure}
\includegraphics[width=7.8cm]{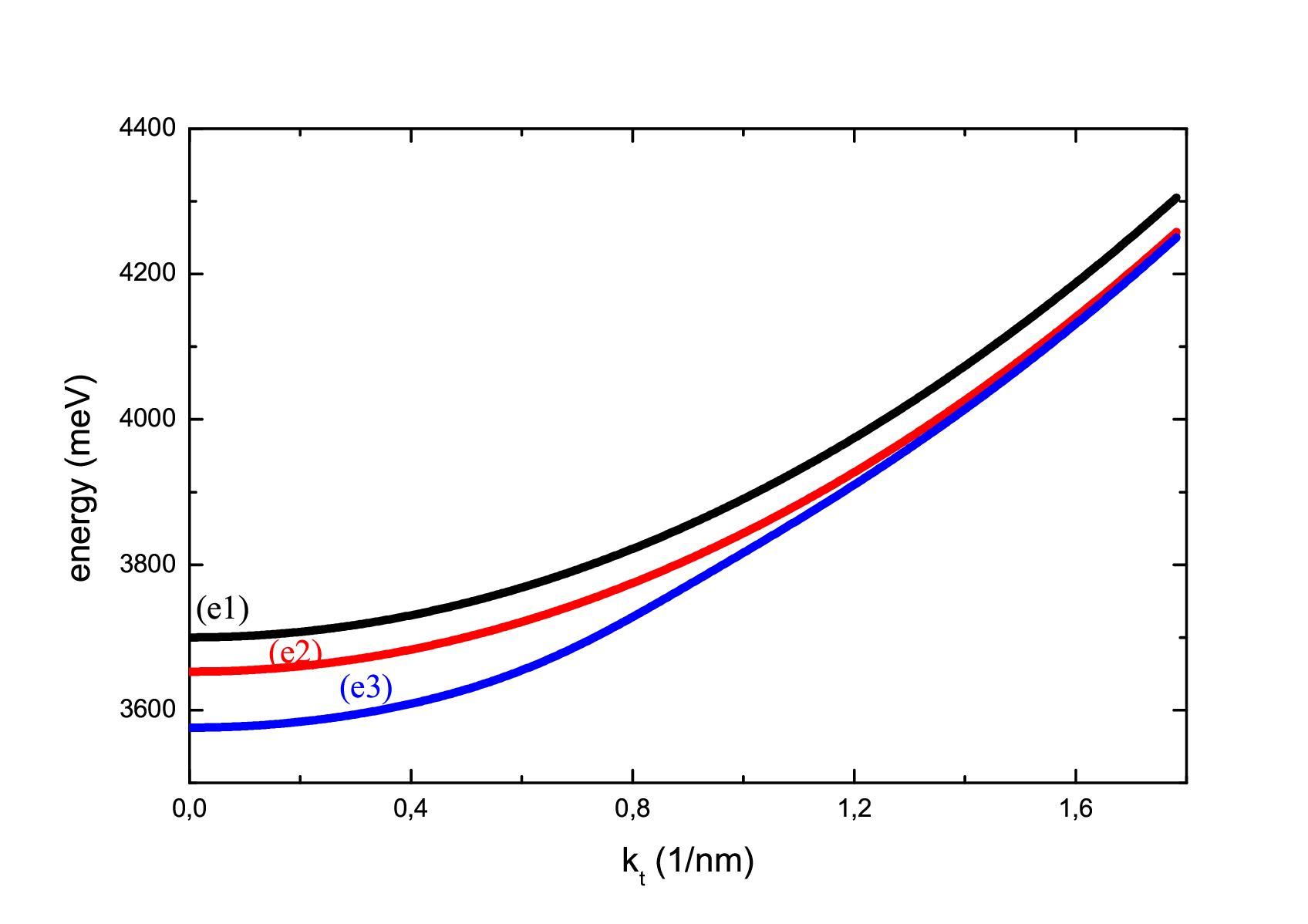}
\vskip-3mm\caption{Calculated energy spectra for electrons (e1) for
the free conduction band, Hartree energy spectra for electrons (e2),
and Hartree--Fock energy spectra for electrons for the quantum well
with a width of 2 nm at the concentration of carriers $n=p=$
$=9\times 10^{12}$ cm$^{-2}$ at the temperature 300~K  }
\end{figure}

The energy spectra for heavy and light holes and for electrons in a
quantum well, as well as the Hartree and
Hartree--Fock renormalizations of the energy spectrum for heavy and
light holes and electrons which reflect the many-body effect known as
a renormalization of the band gap, are presented in Figs. 6 and 7.

\section{Summary}

The calculations of light gain spectra and exciton spectra were
previously carried out only for the nitride quantum well with
parabolic bands and not for quantum wells with compound bands. Here,
we study the effect of nonparabolicity on exciton states in the
wurtzite quantum well. We have calculated and explained that the
exciton binding energy strongly depends on the mixing of valence
bands, because it depends on the overlap integral of the electron
and hole wave functions. We have calculated and explained a shift of
the exciton resonance, which depends on the electron-hole gas
concentration, and the gain spectrum shape in the wurtzite quantum
well. We have found the exchange renormalization of the energy
spectrum for holes and electrons. In the research of the influence
of the overlap integral of wave functions on the Hartree--Fock
renormalization of the electric dipole moment in the wurtzite
quantum well, we conclude that a deviation from a parabolic band
structure in the wurtzite quantum well leads to significant changes
in the determination of the exciton binding energy. The calculations
testify to a small change of the overlap integral of the electron
and hole wave functions, which is caused by the intrinsic quantum
confined Stark effect at the considered concentrations. The
deviation from a parabolic band structure of the quantum well leads
also to significant changes in the overlap integral of the electron
and hole wave functions. This is the cause for a red shift of the
exciton resonance with increasing the concentrations. The
above-presented results can be explained by the influence of the
valence band structure on quantum confined
\mbox{effects.}

\vskip2mm The author is grateful to Prof. V.A.~Kochelap for numerous
discussions.

\vspace*{-5mm}
\rezume{Л.O. Локоть}{%
ХАРТРІ--ФОКІВСЬКА ЗАДАЧА\\ ЕЛЕКТРОННО-ДІРКОВОЇ ПАРИ\\ В КВАНТОВІЙ
ЯМІ GaN }  {Розглянуто мікроскопічне обчислення спектра поглинання
для системи
$\textrm{GaN}/\textrm{Al}_{x}\textrm{Ga}_{1-x}\textrm{N}$ квантової
ями. Тоді як структури квантової ями з параболічним законом
дисперсії проявляють звичайне висвітлювання екситону без зміни
спект\-раль\-ної області, то значне червоне зміщення екситонного
резо\-нансу знайдено для в'юрцитної кван\-то\-воям\-ної
струк\-ту\-ри. Обчис\-ле\-но енергію екситонного резонансу для
в'юр\-цит\-ної квантової ями. Одержані результати можуть
поясню\-ватися впливом ва\-лент\-ної зонної структури на ефекти
квантового конфайнменту. Обчислено оптичний спектр підсилення в
хартрі--фоківській апроксимації. Обчислено зоммерфельдівське
підсилення. Обчис\-ле\-но чер\-воне змі\-ще\-ння спект\-ра
під\-си\-ле\-ння в хартрі--\linebreak фоківській апроксимації
відносно хартрівського спектра підсилення.}

\end{document}